\documentclass[aps,prl,twocolumn,superscriptaddress]{revtex4}
\usepackage{graphicx}
\usepackage{color}

\begin{document}
\newcommand{\tcr}[1]{\textcolor{red}{#1}}


\title{Spin noise spectroscopy in GaAs (110) quantum wells:\\ Access to intrinsic spin lifetimes and equilibrium electron dynamics}


\author{Georg M. M\"uller}
\email[Electronic mail: ]{mueller@nano.uni-hannover.de}
\author{Michael R\"omer}
\affiliation{Institut f\"ur Festk\"orperphysik, Leibniz Universit\"at Hannover,
Appelstra\ss{}e 2, 30167 Hannover, Germany}

\author{Dieter Schuh}
\author{Werner Wegscheider}
\affiliation{Institut f\"ur Experimentelle und Angewandte Physik, Universit\"at Regensburg, 93040 Regensburg, Germany}

\author{Jens H\"ubner}
\author{Michael Oestreich}
\affiliation{Institut f\"ur Festk\"orperphysik, Leibniz Universit\"at Hannover, Appelstra\ss{}e 2, 30167 Hannover, Germany}


\date{\today}

\begin{abstract}
In this letter, the first spin noise spectroscopy measurements in
semiconductor systems of reduced effective dimensionality are
reported. The non-demolition measurement technique gives access to
the otherwise concealed intrinsic, low temperature electron spin
relaxation time of n-doped  GaAs (110) quantum wells and to the
corresponding low temperature anisotropic spin relaxation. The
Brownian motion of the electrons within the spin noise probe laser
spot becomes manifest in a modification of the spin noise line
width. Thereby, the spatially resolved observation of the
stochastic spin polarization uniquely allows to study electron
dynamics at equilibrium conditions with a vanishing total momentum
of the electron system.
\end{abstract}


\maketitle %

The dream of spin quantum information processing and spin based optoelectronic
devices drives the current intense research on spin physics in semiconductors.
Especially GaAs quantum wells (QW) with their quantization axis oriented in
(110)-direction attract great attention since the electron spin relaxation
times in these structures are extremely long even at room temperature. In bulk
semiconductors with zinc blende structure, like GaAs, the spin dephasing of
electrons is dominated over a wide temperature and doping range by the
Dyakonov-Perel (DP) mechanism \cite{dyakonov:spss:13:3023:1972}: The lack of
crystal inversion symmetry leads to a precession of the electron spin in the
wave vector dependent Dresselhaus field $\mathbf{B}(\mathbf{k})$
\cite{dresselhaus:pr:100:580:1955}. In (110) oriented GaAs QWs, the in-plane
components of $\mathbf{B}(\mathbf{k})$ vanish to all orders in the quasi
momentum due to the special crystallographic symmetry and the quantum
confinement in growth direction \cite{winkler:prb:69:045317:2004}. Thereby, the
lifetime $\tau_{z}$ of electron spins aligned in growth direction of the (110) QW
 is considerably larger than in a (100) QW as experimentally shown first
by Ohno \textit{et al.} \cite{ohno:prl:83:4196:1999}. The spin relaxation time is not
infinite but limited at high temperatures by intersubband electron scattering
induced spin relaxation (ISR) \cite{dohrmann:prl:93:147405:2004}. At low
temperatures, the efficiency of ISR is negligible but the well known Bir Aronov
Pikus (BAP) mechanism obviates in nearly all photoluminescence based
experiments the observation of long $\tau_{z}$. The BAP mechanism is caused
by the spin interaction of the photo-created holes with the electrons in the
conduction band which increases with decreasing temperature, i.e.,
photoluminescence (PL) measurements yield shorter $\tau_{z}$ with decreasing
temperature. The BAP mechanism can be avoided in (110) QWs by spatially
separating electron and holes by surface acoustic waves
\cite{couto:prl:98:036603:2007}. However, the influence of the surface acoustic
waves on $\tau_{z}$ is not clear yet. In other words, the intrinsic,
undisturbed $\tau_{z}$ at low temperatures is unknown in (110) QWs.

In this letter, we will show that the intrinsic, low
temperature $\tau_{z}$ in modulation n-doped GaAs (110)  QWs can be measured
by the non-demolition measurement technique of spin noise spectroscopy and that
the intrinsic $\tau_{z}$ is by more than one order of magnitude longer than
measured by Ohno \textit{et al.} \cite{ohno:prl:83:4196:1999} and D\"ohrmann
\textit{et al.} \cite{dohrmann:prl:93:147405:2004} by time resolved PL.
%
Spin noise spectroscopy (SNS) is a measurement technique known in quantum
optics \cite{crooker:nature:431:49:2004} and has been transferred
 to semiconductor systems just recently
\cite{oestreich:prl:95:216603:2005,romer:rsi:78:103903:2007}. In SNS, the
statistical fluctuations of the spin polarization are mapped via Faraday
rotation onto the light polarization of a linear polarized, continuous-wave
laser. The temporal dynamics of the  spin fluctuations  are characterized  by  the electron spin lifetime $\tau$
and, additionally, in the case of a magnetic field in Voigt geometry, by the  precessional 
frequency $\omega$. In the frequency domain, these temporal spin
fluctuations translate to a Lorentzian line shape centered at
$\omega/2\pi$ with a full width at half maximum of $1/\pi \tau$
\cite{koenig:prb:75:085310:2007,romer:rsi:78:103903:2007}.
In SNS experiments, the energy of the probing laser light is
usually chosen  to lie well below the energy gap where excitation
of the semiconductor system is
negligible \cite{romer:rsi:78:103903:2007}. Therefore, SNS allows 
experimental access to the electron spin dynamics near equilibrium
without evoking parasitic spin relaxation by BAP.

The light source for the SNS measurements is  a low noise, tunable diode laser
in Littman configuration \cite{littman:applopt:17:2224:1978}. A Faraday
isolator avoids disturbing feedback and a spatial filter ensures a Gaussian
beam profile. The laser light is focused to a beam waist of $w\approx
3.5\,\mathrm{\mu m}$ on the sample which is mounted in a He cold finger
cryostat. Magnetic fields up to $\mu_0H_{\mathrm{ext}}=14\,\mathrm{mT}$ can be
applied in Voigt geometry. The transmitted light is recollimated  and the rotation of the linear light polarization is resolved by a
combination of a polarizing beam splitter and a high bandwidth balanced photo receiver. The detected electrical signal is amplified by a low
noise amplifier (40 dB) and sent through a low pass filter (-3dB at
67 MHz). The fluctuation signal is digitized with 180~MHZ in the time domain and Fourier
transformed in real time.

In previous SNS measurements on bulk GaAs, the spin noise signal
is shifted from zero frequency
by applying a magnetic field in Voigt geometry. In (110)-oriented GaAs QWs,
this method becomes difficult since spin relaxation is dominated in (110) QWs
at finite magnetic fields by the DP dominated spin relaxation of the in-plane
spin component.
Therefore, small applied magnetic fields   in
Voigt geometry only lead to a broadening and, consequently, the spin noise spectra in this work  are all
centered around zero frequency. For the majority of spectra, large spurious
background noise is most reliably removed by a double difference method: a) The
polarization of the laser light is switched between linear and circular
polarization by a motorized Soleil Babinet compensator in front of the sample.
Circular polarized light is not sensitive to changes in the
Faraday rotation angle and, therefore, contains no spin noise information. b)
For each polarization the applied magnetic field is changed between
$\mu_0H_{\mathrm{ext}}=14\,\mathrm{mT}$, at which the spin noise power becomes
negligible in the observed frequency span, and zero or small magnetic fields.
Subsequently, the noise power spectra for linear and circular polarized light
are subtracted from each other, once for the applied magnetic field of
$\mu_0H_{\mathrm{ext}}=14\,\mathrm{mT}$ and once for a vanishing or smaller
magnetic field. These two curves are on their part again subtracted from each
other and the result is divided by a background noise spectrum  to account for frequency
dependent amplification.

The investigated sample consists of ten identical, symmetrically grown,
nominally 16.8~nm thick GaAs QWs separated by 80~nm Al$_{0.39}$Ga$_{0.61}$As
barriers grown by molecular beam epitaxy and separated from the undoped (110)
GaAs substrate by a 150~nm Al$_{0.80}$Ga$_{0.2}$As lift-off sacrifice layer.
The QWs are symmetrically modulation doped by Si $\delta$-layers in the middle
of the barriers and transport measurements under illumination yield a doping
sheet density of $n=1.8\cdot 10^{11}\, \mathrm{cm^{-2}}$ at 1.5~K.
Photogalvanic experiments by Belkov \textit{et al.} \cite{belkov:prl:100:176806:2008}
have shown that spin dephasing due to structure inversion asymmetry is minimal
in such symmetrically doped (110) QWs in contrast to symmetrically doped (100)
QWs. The substrate is removed for the transmission SNS measurements following
the lift-off recipe of Yablonovitch \textit{et al.}
\cite{yablonovitch:apl:56:2419:1990}, and the multi QW layer is van der Waals
bonded to a c-cut Sapphire substrate. White light transmission measurements
identify the optical absorption edge (interband transition to the Fermi level
energy) of the QWs between 813 and 814~nm.

\begin{figure}
   
        \includegraphics{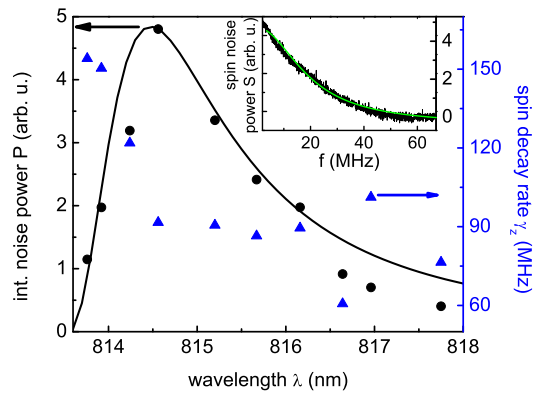}
    \caption{(Color online) Spin decay rate $\gamma_z = 1/\tau_z$ 
    and relative
    integrated spin noise power  at $T=20$~K as a function the laser
    wavelength. Solid line: Integrated spin noise power according
    to model \cite{romer:rsi:78:103903:2007} (details in text).
    The inset shows a typical spin noise spectrum.}
    \label{fig:FigWavelength}
\end{figure}

The inset  of Fig.~\ref{fig:FigWavelength} shows a typical SNS
spectrum measured at a laser wavelength of about 814.25~nm and a temperature of 20~K.
The measured spin noise spectra are fitted with a Lorentz function centered at
zero frequency. The area under the Lorentz curve gives the integrated spin
noise power and the width determines the spin lifetime or spin decay rate
$\gamma$, respectively. Figure~\ref{fig:FigWavelength} shows both the measured
integrated spin noise power (black dots) and the spin relaxation rate (blue
triangles) in dependence on laser wavelength for a lattice temperature of 20~K.
The spin relaxation rate increases sharply when the laser wavelength approaches
the optical absorption edge. This observation is consistent with the fact that
in this temperature regime traditional spin dephasing measurements based on
optical excitation yield results which are completely obstructed by the BAP
mechanism. 
Measurements of this increase of $\gamma_z$ by SNS for wavelengths shorter than
813.7~nm are hindered by the fact that the integrated spin noise power
approaches zero at the optical absorption edge and by the fixed electrical
frequency bandwidth of the detection setup. For wavelengths longer than 815~nm,
optical absorption becomes negligible and the measured spin decay rate is in
good approximation constant indicating that the residual spin life time is
determined by other spin relaxation mechanisms than BAP. Further experiments
presented below show that this spin decay rate is not yet the intrinsic spin
relaxation time but limited by time of flight broadening.

\begin{figure}
    \centering
        \includegraphics{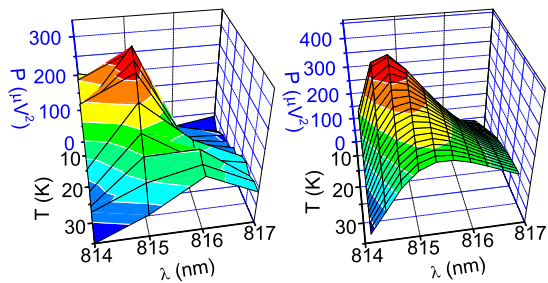}
    \caption{(Color online) Absolute integrated Spin noise power as a
    function of laser wavelength and temperature, measured
    (left panel) and calculated (right panel, details in text).
    }
    \label{fig:FigTemp}
\end{figure}

The black dots and and the solid black line in Fig.~\ref{fig:FigWavelength}
depict the measured and the calculated integrated spin noise power,
respectively. The integrated spin noise power $P$ is calculated by a
phenomenological model which is based on the change of the refractive index due
to the fluctuating imbalance of electron spins at the Fermi energy
\cite{romer:rsi:78:103903:2007}: $ P\propto (\left. d n/d \alpha
\right|_{\alpha_0}  \cdot N_{\mathrm{fluc}}/N_{\mathrm{bleach}}\,\alpha_0)^2$,
where $n$ is the real part of the refractive index, $N_{\mathrm{fluc}}$ the
root mean square imbalance between the two spin directions,
$N_{\mathrm{bleach}}$ the critical imbalance that would bleach the optical
transition for one spin direction, and $\alpha$ the absorption constant. The
transition is set to 813.5\,nm in accordance with the optical transmission
measurements. The calculated wavelength dependence with the minimum of $P$
at the optical absorption edge agrees perfectly with the experimental data. Figure~\ref{fig:FigTemp} shows a more detailed comparison of the
measured and calculated $P$ in dependence on temperature and
wavelength. The following sample parameters are used for the calculation: The polarizability due to the optical selection rules of the
optical transition is set to $\beta=0.6$ \cite{pfalz:prb:71:165305:2005} and an optical
absorption constant $\alpha_0=2.28\cdot 10 ^6 \,\mathrm{m}^{-1}$ is assumed \footnote{The
optical absorption constant is not measured by an independent experiment but
has been adjusted in the SNS calculations to the measured integrated spin noise
power at 815~nm and 20~K in Fig. \ref{fig:FigTemp}. Nonetheless, the
adjusted $\alpha_0$ is in good agreement with typical absorption measurements
on similar QWs.}.  A line width of the optical resonance of
$k_{\mathrm{B}}\cdot 20$~K for $T\leq20$~K and of
$k_{\mathrm{B}}\cdot T$ for $T>20$~K is used within the
calculations since the width of the measured white light transmission
change is constant below 20~K and increases linearly above. The  energy of the resonance shifts with the temperature dependent Fermi level energy.  Also
this two-dimensional dependence of the calculated SNS on
wavelength and temperature is in good agreement with our
measurements, demonstrating that the origin of the spin noise
signal is well understood.

\begin{figure}
    \centering
        \includegraphics{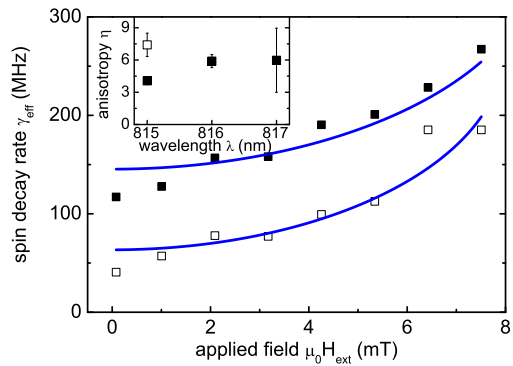}
    \caption{(Color online) Effective spin decay rate
    $\gamma_{\mathrm{eff}} = 1/\tau_{\mathrm{eff}}$ ($T=20$ K) as a function of the applied
    magnetic field with fits according to Eq. (\ref{EqAniso}).
    Inset: Anisotropy determined by the fits as function of
    laser wavelength ($T=20$ K). Open symbols indicate
    measurements with an enlarged focus in which time of
    flight broadening is strongly reduced.}
    \label{fig:FigField}
\end{figure}

Next, we observe the influence of the spin relaxation anisotropy on the spin
noise signal, which has previously been examined at relatively high magnetic
fields \cite{dohrmann:prl:93:147405:2004}, where the spin relaxation anisotropy
slows down the effective Larmor frequency. However for very small magnetic
fields, with $\omega_{\mathrm{L}} <
(\gamma_{\mathrm{\bot}}-\gamma_{\mathrm{z}})/2$, there is no
precessional motion of the spins as the spin decays due to the large in-plane
spin relaxation rate $\gamma_{\mathrm{\bot}}$ before even half a rotation is
carried out. The effective spin relaxation rate is in this case
\begin{equation}
\label{EqAniso}
\gamma_{\mathrm{eff}}=\frac{\gamma_{\mathrm{\bot}}+\gamma_{\mathrm{z}}}{2}+
\frac{\gamma_{\mathrm{\bot}}-\gamma_{\mathrm{z}}}{2}\sqrt{1-4\frac{\omega_{\mathrm{L}}^2}{(\gamma_{\mathrm{\bot}}-\gamma_{\mathrm{z}})^2}}.
\end{equation}
Figure~\ref{fig:FigField} depicts the measured $\gamma_{\mathrm{eff}}$ (black
squares) as a function of applied magnetic field at T=20~K. We have measured
$\omega_L$, i.e., an electron g-factor $g^\ast=0.29$, on the same sample by time-resolved PL at B=4~T and fitted the spin noise data by a least square fit
corresponding to Eq.~\ref{EqAniso} (top blue solid curve). The fit directly
yields the anisotropy factors $\eta=\gamma_{\mathrm{\bot}}/\gamma_{\mathrm{z}}$
which is shown as filled squares in the inset of Fig.~\ref{fig:FigField} in
dependence on the laser wavelength. The anisotropy factor is at 815~nm smaller
than at longer wavelengths since BAP is in contrast to DP at most weakly
dependent on the crystallographic direction, i.e., an efficient isotropic spin
relaxation lowers the spin relaxation anisotropy. In agreement with the
wavelength dependent data (Fig.~\ref{fig:FigWavelength}), the anisotropy factor
is constant for long wavelengths since BAP is switched off.

Anisotropy measurements at 815~nm with a defocused, i.e., enlarged, laser spot
on the sample reduce the spin relaxation by BAP and yield an anisotropy factor
$\eta=7.4(1.0)$ which is a almost factor of two larger than in the focused case.  
This anisotropy factor is
of the same magnitude as $\eta$ measured in a similar sample at room
temperature \cite{dohrmann:prl:93:147405:2004}  and in an undoped GaAs (110) QW at low temperature \cite{couto:prl:98:036603:2007}. However, the physical origins are different. In the room temperature case, the anisotropy is limited by ISR and, in the latter case, the anisotropy is probably dominated by the yet unclear influence of the surface acoustic waves. As it will be argued later in this letter, the anisotropy measured in this work is in contrast given by the intrinsic low temperature spin lifetimes of the sample.  
Still, the strong measured increase of $\tau_{z}$ after defocusing can not be solely explained by
the reduction of BAP since $\eta$ is only equal to about 6 for laser wavelengths of
816 and 817~nm where BAP is negligible (see Fig.~\ref{fig:FigField}). In fact,
the strong dependence on the laser spot diameter indicates diffusion of
electrons out of the laser spot which is equivalent to time of flight
broadening.
\begin{figure}
    \centering
        \includegraphics{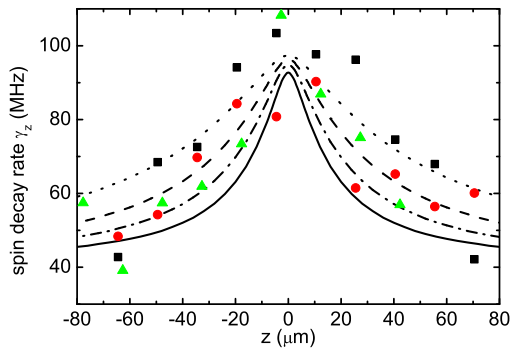}
    \caption{(Color online) Effective Spin decay rate $\gamma_z = 1/\tau_z$
    ($T=20$ K) as a function of sample position $z$ for  a laser wavelength
    of $\lambda = 815$ nm (squares), $\lambda = 816$ nm (circles) and
    $\lambda = 816$ nm with doubled laser power (triangles). The $z=0$
    position is set to the maxima of the measured curves. The lines show
    calculations according to the model given by Eq. (\ref{EqTof}) for
    $D_{\mathrm{eff}}=100\,\mathrm{cm^2/s}$ (straight line), $200\,\mathrm{cm^2/s}$
    (dot and dash  line), $400\,\mathrm{cm^2/s}$ (dashed line) and $1000\,\mathrm{cm^2/s}$ (dotted line).}
    \label{fig:FigDefocus}
\end{figure}
Figure~\ref{fig:FigDefocus} depicts the measured spin decay rate in dependence
on the defocusing distance $z$ which is the distance between the focus of the
Gaussian laser beam and the sample. The filled squares show $\gamma_z$ measured
at 815~nm where BAP can not be neglected and the filled circles show $\gamma_z$
measured at 816~nm where BAP is unimportant. Mesurements at 816~nm with twice
the laser power (filled triangles) rule out that the observations can be completely attributed to
excitation density dependent spin dephasing mechanisms as BAP.
To model this time of flight broadening, we extend the existing
spin noise model \cite{romer:rsi:78:103903:2007} by taking into
account the classical position of the electron within the laser
beam:
\begin{equation}
\label{EqTof} S(\omega)=\int d \mathbf{r}_{\mathrm{0}} \left|
\mathcal{F} \left\{ \int  d \mathbf{r} \exp(-\gamma_z^{\mathrm{intr}} t) \cdot
P(\mathbf{r}, \mathbf{r}_{\mathrm{0}})\cdot
I(\mathbf{r})\right\}\right|^2.
\end{equation}
Hereby, the two dimensional vector $\mathbf{r}_{\mathrm{0}}$ gives the position
of an individual electron at the beginning of the measurement, $\exp(-\gamma_z^{\mathrm{intr}}
t)$ describes the intrinsic spin decay, $I(\mathbf{r})=I_0\exp(-2r^2/w(z)^2)$
weights the position of the electron in the laser spot, and $P(\mathbf{r},
\mathbf{r}_{\mathrm{0}})$ gives the probability distribution for classical
Brownian motion in two dimensions with respect to a diffusion constant $D$
\cite{bergmann:prb:28:2914:1983}: $P(\mathbf{r},
\mathbf{r}_{\mathrm{0}})=1/4\pi D\, t\cdot \exp(-(\mathbf{r}-
\mathbf{r}_{\mathrm{0}})^2/4 D\, t)$.
The dependence of the effective spin decay rate on the sample position $z$ is
calculated numerically  with Eq. (\ref{EqTof}) and is plotted in Fig.
\ref{fig:FigDefocus} for different values of $D$. The experimentally determined
beam parameters are: waist  $w_0=3.5\,\mathrm{\mu m}$ and  Rayleigh range
$z_{\mathrm{R}}=3\,\mathrm{\mu m}$. As intrinsic spin decay rate, we determine
$\gamma_{z}^{\mathrm{intr}} = 42 (2.5)\,\mathrm{MHz}$. This value is the average decay rate
to which the measured values converge when strongly defocused
($T=20\,\mathrm{K}$ and $\lambda=815\,\mathrm{nm}$) and corresponds to a spin
lifetime of $\tau_{z}^{\mathrm{intr}} = 24(2)$\,ns which is to our knowledge the longest
measured spin lifetime in  n-doped GaAs (110) QWs.  Best agreement between model and
experiment is obtained for values of $D$ between $100$ and
$1000\,\mathrm{cm^2/s}$ (see Fig. \ref{fig:FigDefocus}). The measured mobility
at low temperature under illumination of $\mu=3.7\cdot 10^5\,\mathrm{cm^2/Vs}$
gives together with the Einstein relation $D=\frac{\mu
E_{\mathrm{F}}}{e}\approx 2400\,\mathrm{ cm^2/s}$
($E_{\mathrm{F}}=6.4\,\mathrm{meV}$). Contrary to electrical mobility
measurements, in our experiment the electronic system has vanishing total
momentum. Therefore, the Brownian motion within the laser spot is governed by
electron-electron collisions that keep the total momentum of the electronic
system constant and only represent a slight modification to the electrical
conductivity that is mainly determined by electron-phonon and electron-impurity
collisions. The electron-electron scattering time in thermal equilibrium
calculates to
$\tau_{\mathrm{ee}}=1/[\frac{\pi}{2}\left(k_{\mathrm{B}}T\right)^2 / \hbar
E_{\mathrm{F}}\ln E_{\mathrm{F}}/k_{\mathrm{B}}T]\approx 700\,\mathrm{fs}$ for
our sample system \cite{fasol:apl:59:2430:1991,fukuyama:prb:27:5976:1983}.
Calculating a diffusion constant  from the Einstein relation together with the
Drude model gives $D_{\mathrm{eff}}=118\,\mathrm{cm^2/s}$. Thus, this coarse
estimate yields an effective diffusion constant of the correct order of
magnitude and shows that the experiment is sensitive enough to study electron
motion near thermal equilibrium. 

We attribute the residual intrinsic spin dephasing rate of $\gamma_{z}^{\mathrm{intr}} =
42(2.5)\,\mathrm{MHz}$ to a DP mechanism due to random Rashba fields caused by
a fluctuating donor density in the symmetric doping layers
\cite{sherman:apl:82:209:2003}. The relevant area on which these donor density
fluctuations between the layers have to be considered is given by $A=\pi
(v_{\mathrm{F}}\tau_{\mathrm{ee}})^{2}$. Assuming a Poisson distribution, the
donor fluctuations are of the order of $1/\sqrt{A\,n}\approx 11\%$. The
resulting electric field is calculated  in a simpe parallel-plate capacitor
model. With the Rashba coefficient and  Eq. (3) of Ref.
\onlinecite{eldridge:prb:77:1253442008} modified for the case of a degenerate
electron gas, we obtain a spin dephasing rate of 10 MHz, which compares well in
view of the approximations made.

In conclusion, we have shown that SNS allows to measure intrinsic spin
lifetimes in n-doped GaAs  (110)  QWs. These lifetimes are limited by intrinsic,
random Rashba fields that stem from unavoidable fluctuations in the donor
density. Furthermore, we have shown that the spin noise spectra are time of flight
broadened and theoretically described by a model which predicts that spin noise
spectroscopy allows to study near equilibrium electron dynamics. Therefore, at
millikelvin temperatures, which are generally accessible
due to the non-perturbative nature of spin noise spectroscopy, modifications to
the classical motion like weak localization effects
\cite{bergmann:prb:28:2914:1983} should become directly accessible at thermal
equilibrium.

This work was supported by the DFG and the BMBF (NanoQuit). The authors thank Stefan Oertel for the measurement of the g-factor. G.M.M.
acknowledges a fellowship by the Evangelisches Studienwerk.


\begin{thebibliography}{19}
\expandafter\ifx\csname natexlab\endcsname\relax\def\natexlab#1{#1}\fi
\expandafter\ifx\csname bibnamefont\endcsname\relax
  \def\bibnamefont#1{#1}\fi
\expandafter\ifx\csname bibfnamefont\endcsname\relax
  \def\bibfnamefont#1{#1}\fi
\expandafter\ifx\csname citenamefont\endcsname\relax
  \def\citenamefont#1{#1}\fi
\expandafter\ifx\csname url\endcsname\relax
  \def\url#1{\texttt{#1}}\fi
\expandafter\ifx\csname urlprefix\endcsname\relax\def\urlprefix{URL }\fi
\providecommand{\bibinfo}[2]{#2}
\providecommand{\eprint}[2][]{\url{#2}}

\bibitem[{\citenamefont{D'yakonov and
  Perel'}(1972)}]{dyakonov:spss:13:3023:1972}
\bibinfo{author}{\bibfnamefont{M.~I.} \bibnamefont{D'yakonov}}
  \bibnamefont{and} \bibinfo{author}{\bibfnamefont{V.~I.}
  \bibnamefont{Perel'}}, \bibinfo{journal}{Sov. Phys. Solid State}
  \textbf{\bibinfo{volume}{13}}, \bibinfo{pages}{3023} (\bibinfo{year}{1972}).

\bibitem[{\citenamefont{Dresselhaus}(1955)}]{dresselhaus:pr:100:580:1955}
\bibinfo{author}{\bibfnamefont{G.}~\bibnamefont{Dresselhaus}},
  \bibinfo{journal}{Phys. Rev.} \textbf{\bibinfo{volume}{100}},
  \bibinfo{pages}{580} (\bibinfo{year}{1955}).

\bibitem[{\citenamefont{Winkler}(2004)}]{winkler:prb:69:045317:2004}
\bibinfo{author}{\bibfnamefont{R.}~\bibnamefont{Winkler}},
  \bibinfo{journal}{Phys. Rev. B} \textbf{\bibinfo{volume}{69}},
  \bibinfo{pages}{045317} (\bibinfo{year}{2004}).

\bibitem[{\citenamefont{Ohno et~al.}(1999)\citenamefont{Ohno, Terauchi, Adachi,
  Matsukura, and Ohno}}]{ohno:prl:83:4196:1999}
\bibinfo{author}{\bibfnamefont{Y.}~\bibnamefont{Ohno}} \textit{et al.},
  \bibinfo{journal}{Phys. Rev. Lett.} \textbf{\bibinfo{volume}{83}},
  \bibinfo{pages}{4196} (\bibinfo{year}{1999}).

\bibitem[{\citenamefont{D\"ohrmann et~al.}(2004)\citenamefont{D\"ohrmann,
  H\"agele, Rudolph, Bichler, Schuh, and
  Oestreich}}]{dohrmann:prl:93:147405:2004}
\bibinfo{author}{\bibfnamefont{S.}~\bibnamefont{D\"ohrmann}} \textit{et al.},
  \bibinfo{journal}{Phys. Rev. Lett.} \textbf{\bibinfo{volume}{93}},
  \bibinfo{eid}{147405} (\bibinfo{year}{2004}).

\bibitem[{\citenamefont{Couto et~al.}(2007)\citenamefont{Couto, Iikawa,
  Rudolph, Hey, and Santos}}]{couto:prl:98:036603:2007}
\bibinfo{author}{\bibfnamefont{O.~D.~D.} \bibnamefont{Couto},
  \bibfnamefont{Jr.}} \textit{et al.},
  \bibinfo{journal}{Phys. Rev. Lett.} \textbf{\bibinfo{volume}{98}},
  \bibinfo{pages}{036603} (\bibinfo{year}{2007}).

\bibitem[{\citenamefont{Crooker et~al.}(2004)\citenamefont{Crooker, Rickel,
  Balatsky, and Smith}}]{crooker:nature:431:49:2004}
\bibinfo{author}{\bibfnamefont{S.~A.} \bibnamefont{Crooker}} \textit{et al.},
  \bibinfo{journal}{Nature} \textbf{\bibinfo{volume}{431}}, \bibinfo{pages}{49}
  (\bibinfo{year}{2004}).

\bibitem[{\citenamefont{Oestreich et~al.}(2005)\citenamefont{Oestreich,
  R\"{o}mer, Haug, and H\"{a}gele}}]{oestreich:prl:95:216603:2005}
\bibinfo{author}{\bibfnamefont{M.}~\bibnamefont{Oestreich}} \textit{et al.},
  \bibinfo{journal}{Phys. Rev. Lett.} \textbf{\bibinfo{volume}{95}},
  \bibinfo{eid}{216603} (\bibinfo{year}{2005}).

\bibitem[{\citenamefont{R\"omer et~al.}(2007)\citenamefont{R\"omer, H\"ubner,
  and Oestreich}}]{romer:rsi:78:103903:2007}
\bibinfo{author}{\bibfnamefont{M.}~\bibnamefont{R\"omer}},
  \bibinfo{author}{\bibfnamefont{J.}~\bibnamefont{H\"ubner}}, \bibnamefont{and}
  \bibinfo{author}{\bibfnamefont{M.}~\bibnamefont{Oestreich}},
  \bibinfo{journal}{Rev. Sci. Instrum.} \textbf{\bibinfo{volume}{78}},
  \bibinfo{pages}{103903} (\bibinfo{year}{2007}).

\bibitem[{\citenamefont{Braun and K\"{o}nig}(2007)}]{koenig:prb:75:085310:2007}
\bibinfo{author}{\bibfnamefont{M.}~\bibnamefont{Braun}} \bibnamefont{and}
  \bibinfo{author}{\bibfnamefont{J.}~\bibnamefont{K\"{o}nig}},
  \bibinfo{journal}{Phys. Rev. B} \textbf{\bibinfo{volume}{75}},
  \bibinfo{eid}{085310} (\bibinfo{year}{2007}).

\bibitem[{\citenamefont{Littman and
  Metcalf}(1978)}]{littman:applopt:17:2224:1978}
\bibinfo{author}{\bibfnamefont{M.~G.} \bibnamefont{Littman}} \bibnamefont{and}
  \bibinfo{author}{\bibfnamefont{H.~J.} \bibnamefont{Metcalf}},
  \bibinfo{journal}{Appl. Opt.} \textbf{\bibinfo{volume}{17}},
  \bibinfo{pages}{2224} (\bibinfo{year}{1978}).

\bibitem[{\citenamefont{Bel'kov et~al.}(2008)\citenamefont{Bel'kov, Olbrich,
  Tarasenko, Schuh, Wegscheider, Korn, Schuller, Weiss, Prettl, and
  Ganichev}}]{belkov:prl:100:176806:2008}
\bibinfo{author}{\bibfnamefont{V.~V.} \bibnamefont{Bel'kov}} \textit{et al.},
  \bibinfo{journal}{Phys. Rev. Lett.} \textbf{\bibinfo{volume}{100}},
  \bibinfo{pages}{176806} (\bibinfo{year}{2008}).

\bibitem[{\citenamefont{Yablonovitch et~al.}(1990)\citenamefont{Yablonovitch,
  Hwang, Gmitter, Florez, and Harbison}}]{yablonovitch:apl:56:2419:1990}
\bibinfo{author}{\bibfnamefont{E.}~\bibnamefont{Yablonovitch}} \textit{et al.}, \bibinfo{journal}{Appl. Phys. Lett.}
  \textbf{\bibinfo{volume}{56}}, \bibinfo{pages}{2419} (\bibinfo{year}{1990}).

\bibitem[{\citenamefont{Pfalz et~al.}(2005)\citenamefont{Pfalz, Winkler,
  Nowitzki, Reuter, Wieck, H\"agele, and Oestreich}}]{pfalz:prb:71:165305:2005}
\bibinfo{author}{\bibfnamefont{S.}~\bibnamefont{Pfalz}} \textit{et al.},
  \bibinfo{journal}{Phys. Rev. B} \textbf{\bibinfo{volume}{71}},
  \bibinfo{eid}{165305} (\bibinfo{year}{2005}).

\bibitem[{\citenamefont{Bergmann}(1983)}]{bergmann:prb:28:2914:1983}
\bibinfo{author}{\bibfnamefont{G.}~\bibnamefont{Bergmann}},
  \bibinfo{journal}{Phys. Rev. B} \textbf{\bibinfo{volume}{28}},
  \bibinfo{pages}{2914} (\bibinfo{year}{1983}).

\bibitem[{\citenamefont{Fasol}(1991)}]{fasol:apl:59:2430:1991}
\bibinfo{author}{\bibfnamefont{G.}~\bibnamefont{Fasol}},
  \bibinfo{journal}{Appl. Phys. Lett.} \textbf{\bibinfo{volume}{59}},
  \bibinfo{pages}{2430} (\bibinfo{year}{1991}).

\bibitem[{\citenamefont{Fukuyama and
  Abrahams}(1983)}]{fukuyama:prb:27:5976:1983}
\bibinfo{author}{\bibfnamefont{H.}~\bibnamefont{Fukuyama}} \bibnamefont{and}
  \bibinfo{author}{\bibfnamefont{E.}~\bibnamefont{Abrahams}},
  \bibinfo{journal}{Phys. Rev. B} \textbf{\bibinfo{volume}{27}},
  \bibinfo{pages}{5976} (\bibinfo{year}{1983}).

\bibitem[{\citenamefont{Sherman}(2003)}]{sherman:apl:82:209:2003}
\bibinfo{author}{\bibfnamefont{E.~Y.} \bibnamefont{Sherman}},
  \bibinfo{journal}{Appl. Phys. Lett.} \textbf{\bibinfo{volume}{82}},
  \bibinfo{pages}{209} (\bibinfo{year}{2003}).

\bibitem[{\citenamefont{Eldridge et~al.}(2008)\citenamefont{Eldridge, Leyland,
  Lagoudakis, Karimov, Henini, Taylor, Phillips, and
  Harley}}]{eldridge:prb:77:1253442008}
\bibinfo{author}{\bibfnamefont{P.~S.} \bibnamefont{Eldridge}} \textit{et al.}, \bibinfo{journal}{Phys. Rev. B}
  \textbf{\bibinfo{volume}{77}}, \bibinfo{pages}{125344}
  (\bibinfo{year}{2008}).

\end{thebibliography}

\end{document}